# Multi-level Spectroscopy of Two-Level Systems Coupled to a dc SQUID Phase Qubit

T. A. Palomaki,[a)] S. K. Dutta,[b)] R. M. Lewis,[c)] A. J. Przybysz, Hanhee Paik,[d)] B. K. Cooper, H. Kwon, J. R. Anderson, C. J. Lobb, and F. C. Wellstood

Joint Quantum Institute and Center for Nanophysics and Advanced Materials
Department of Physics, University of Maryland, College Park MD, 20742

and

E. Tiesinga
Joint Quantum Institute
National Institute of Standards and Technology, Gaithersburg, Maryland  20899-8423

## Abstract

We report spectroscopic measurements of discrete two-level systems (TLSs) coupled to a dc SQUID phase qubit with a 16 µm$^2$ area Al/AlO$_x$/Al junction. Applying microwaves in the 10 GHz to 11 GHz range, we found eight avoided level crossings with splitting sizes from 10 MHz to 200 MHz and spectroscopic lifetimes from 4 ns to 160 ns. Assuming the transitions are from the ground state of the composite system to an excited state of the qubit or an excited state of one of the TLS states, we fit the location and spectral width to get the energy levels, splitting sizes and spectroscopic coherence times of the phase qubit and TLSs. The distribution of splittings is consistent with non-interacting individual charged ions tunneling between random locations in the tunnel barrier and the distribution of lifetimes is consistent with the AlO$_x$ in the junction barrier having a frequency-independent loss tangent. To check that the charge of each TLS couples independently to the voltage across the junction, we also measured the spectrum in the 20-22 GHz range and found tilted avoided level crossings due to the second excited state of the junction and states in which both the junction and a TLS were excited.

a) Present address: Department of Microelectronics and Nanoscience, MC2, Chalmers University of Technology, S-41296 Gothenburg, Sweden: electronic address tauno@chalmers.se

b) Present address: Department of Physics, Georgetown University, 37[th] and O St. N.W. Washington, DC 20057

c) Present address: Northrop Grumman,1212 Winterson Road, Linthicum, MD 21090

d) Present address: Department of Applied Physics, Yale University, New Haven, CT 06520

**Introduction**

Dielectric loss and individual spurious charged two-level systems (TLSs) have proven to be major causes of energy relaxation and decoherence in Josephson junction phase qubits [1-3]. Although dielectric loss and TLSs produce distinct effects in a phase qubit, they are merely different manifestations of the motion of defect charges in insulating solids [4-8]. Indeed, in the standard model of dielectric loss in solids, energy relaxation is due to a collection of charged two-level systems formed by ions moving randomly in atomic-scale two-well potentials. Better understanding of the sources of dielectric loss at low temperature and the nature of the spurious two-level systems is needed to improve the performance of phase qubits for quantum computing and to determine if the two-level systems can themselves be exploited as naturally occurring qubits [9].

There are two main ways [10] that a moving ion could couple energetically to a phase qubit and produce an avoided level crossing [11-16]:

(i) *Charge fluctuator:* an ionic charge $Q$ in the dielectric tunnel barrier of a Josephson junction, or in a surrounding dielectric layer, could couple to the voltage $V$ across the junction's capacitance $C$.

(ii) *Critical current fluctuator:* an ion moving in the dielectric tunnel barrier of a Josephson junction could modulate the critical current $I_0$ of the junction.

There are many other types of fluctuations that could lead to decoherence in a qubit, including fluctuations in bias current [17], magnetic flux noise [18-20], and fluctuations in capacitance or inductance. Microwave circuit resonances or cavity modes could couple to a phase qubit and produce avoided level crossings [21]. Another possibility is magnetic spin coupling to the magnetic field produced by the current flowing through the junction. However, we expect this term to be quite small for typical devices. For a single-electron spin, the magnetic fields generated by the microamp currents flowing through our micron scale thin-film wires would be far too small to produce the avoided level crossings seen. We do not consider the behavior of these other mechanisms here, but note that the splittings or decoherence they produce will generally differ in distinguishable ways from that produced by charge and critical current fluctuators.

In general, one would expect a moving ion in the tunnel barrier of a junction to produce both critical current and charge effects simultaneously, with the relative importance depending on microscopic details. Charge and critical current fluctuators produce some distinguishing features. For example, for a charge fluctuator the coupling to the junction scales as $Q/\sqrt{C}$ and the strongest coupling occurs for ions in the junction's tunnel barrier, where the electric field from the junction voltage is most intense, while the coupling is weaker for ions in surrounding insulating layers [2, 22, 23]. For a critical current fluctuator, only ions that are in the tunnel barrier will be coupled and the coupling strength will tend to scale linearly with the critical current density [12, 23]. Comparing the distribution of splitting sizes in devices with different capacitances or critical current densities can thus provide useful insight into the nature of the coupling. However, such statistical characterizations do not allow definitive identification of the coupling mechanism for an individual TLS or the clear identification of fluctuators that have both charge and critical current coupling.

Spectroscopic and time-resolved quantum state measurements of phase qubits provide sensitive new techniques for investigating the properties of individual charged two-level systems and procedures to identify the coupling mechanism have been proposed [13, 14]. However, to our knowledge no concrete experimental conclusions have been reached on whether the TLSs seen in phase qubits are charge or critical current fluctuators. In contrast, recent results on TLSs in charge qubits strongly support a charge coupling model [24].

In this paper we present detailed multi-level spectroscopic measurements of individual TLSs coupled to an Al/AlO$_x$/Al dc SQUID (Superconducting Quantum Interference Device) phase qubit at milli-Kelvin temperatures. Comparing the spectra to charge and critical current coupling models, we test some key aspects of junction-TLS coupling, such as whether the TLSs are coupled to each other. As with other recent work on phase qubits, our results do not uniquely pin down the coupling mechanism. However, we find overall that the data supports the ionic nature of the TLSs and furthermore that the properties of the individual TLSs appear to be distributed in just the form needed to produce a frequency-independent loss tangent. We also examine in detail the coupling between the qubit and the TLS through spectroscopic widths.

**Hamiltonian of a phase qubit and two-level systems**

The Hamiltonian of an isolated phase qubit can be written as [25-27]:

$$H_j(p_\gamma, \gamma) = \left(\frac{2\pi}{\Phi_0}\right)^2 \frac{p_\gamma^2}{2C} - \frac{\Phi_0}{2\pi}(I_0 \cos\gamma + I\gamma) , \qquad (1)$$

where $C$ is the junction capacitance, $I_0$ is the critical current, $I$ is the bias current, $p_\gamma$ is the canonical momentum conjugate to the junction phase difference $\gamma$, and $\Phi_0$ is the flux quantum. Here we will consider a composite system consisting of a phase qubit that is coupled to $N$ two-level systems. Assuming that the two-level systems only interact with the phase qubit and not directly with each other, the Hamiltonian of the combined system can be written as:

$$H = H_j(p_\gamma, \gamma) + \sum_{i=1}^{N} H_i(\bar{p}_i, \bar{x}_i) + \sum_{i=1}^{N} H_{ci} , \qquad (2)$$

where $H_i(\bar{p}_i, \bar{x}_i)$ is the Hamiltonian of the i-th ion, $H_{ci}$ is the coupling Hamiltonian between the i-th ion and the qubit, $\bar{p}_i$ is the momentum operator of the i-th ion and $\bar{x}_i$ is the position of the i-th ion.

To proceed, we choose the basis states of the combined system to be tensor products of the energy eigenstates of the uncoupled system. We initially keep only the lowest two levels of the qubit junction and consider the $2N+2$ basis states of the form $|0g\rangle$, $|1g\rangle$, $|0e_i\rangle$, and $|1e_i\rangle$, corresponding respectively to the junction and all of the ions being in the ground state, the junction being excited and all the ions being in their ground state, the junction being in the ground state and just the i-th ion being in its excited state, and the junction being in its excited state and just the i-th ion also being in its excited state. We define the ground state energy of the ions and the junction to be zero, the energy of the excited state of the i-th ion to be $E_i$ and the energy of the excited state of the junction to be $E_\gamma$.

We now consider the situation where the $n=1$ level of the junction crosses the excited state of the i-th ion, i.e. where $|0e_i\rangle$ and $|1g_i\rangle$ cross. Considering just these two unperturbed states, we can write a reduced Schrodinger's equation in matrix form as:

$$\begin{pmatrix} E_i & \frac{\Delta_i}{2} \\ \frac{\Delta_i^*}{2} & E_\gamma \end{pmatrix} \begin{pmatrix} a \\ b \end{pmatrix} = E \begin{pmatrix} a \\ b \end{pmatrix}, \qquad (3)$$

where E is the energy, $a$ and $b$ are the amplitudes to be in the states $|0e_i\rangle$ and $|1g_i\rangle$, respectively and we have defined the energy splitting term:

$$\Delta_i = 2 < 0e_i | \sum_{i=1}^{N} H_{ci}(p_\gamma, \bar{x}_i) | 1g_i >. \qquad (4)$$

In arriving at Eq. (3), we have also assumed that $< 0e_i | \sum_{i=1}^{N} H_{ci}(p_\gamma, \bar{x}_i) | 0e_i >$ and $< 1g | \sum_{i=1}^{N} H_{ci}(p_\gamma, \bar{x}_i) | 1g >$ vanish; the contribution is small unless the junction is biased into a very anharmonic regime, as shown below.

Solving Eq. (3) for the energy eigenvalues $E$ gives the classic avoided crossing spectrum,

$$E = (E_\gamma + E_i)/2 \pm \sqrt{(E_\gamma - E_i)^2 + |\Delta_i|^2}/2, \qquad (5)$$

with the minimum energy difference $\Delta E = E_+ - E_- = |\Delta_i|$ occurring at the avoided crossing where $E_\gamma = E_i$. Since we have included only one ion in the analysis, Eq. (5) is approximate. Although we have assumed the ions do not directly couple to each other, since each ion couples to the junction, there is an indirect interaction that perturbs ions that have nearly the same energy. For the basis states $|0g\rangle$, $|1g\rangle$, $|0e_i\rangle$, and $|1e_i\rangle$, with $i = 1$ to $N$, accurately determining the energy eigenvalues of the coupled system will require diagonalizing a matrix of size $(2N+2)^2$.

The lifetime of the coupled eigenstates should reflect the lifetimes of the individual quantum systems as well as the coupling among them. This behavior can be modeled by assigning imaginary terms to the uncoupled eigenstate energies and then solving the resulting complex version of Eq. (5). The real component of the resulting complex eigenvalue E is the energy of the state and the imaginary component is equal to $h\Delta f/2$, where $\Delta f$ is the full-width at half-max of the resonance. The spectroscopic coherence time is given by $T_2^* = 1/\pi\Delta f$ and provides an easy way to measure a lower bound on the coherence time $T_2$ of the system since $T_2^* < T_2$ [28].

**Splitting due to charge fluctuators**

For a charge fluctuator, the coupling Hamiltonian between the phase qubit and the i-th ion can be written as

$$H_{ci}(\bar{p}_\gamma, \bar{x}_i) = -Q_i \int_{\bar{x}_o}^{\bar{x}_i} \bar{E}(\bar{r}_i) \cdot d\bar{r}_i, \tag{6}$$

where $\bar{x}_i$ is the position of the i-th ion, $Q_i$ is the charge of the i-th ion, $\bar{E}(\bar{r}_i)$ is the electric field at $\bar{r}_i$, $\bar{x}_o$ is a reference point on the ground plate of the capacitor, and the integral is taken over any path from $\bar{x}_o$ to $\bar{x}_i$. If the ion is located in the junction capacitor C and this capacitor has a parallel plate configuration, the coupling Hamiltonian reduces to:

$$H_{ci}(\bar{p}_\gamma, \bar{x}_i) = \frac{2\pi}{Cd\Phi_o} Q_i p_\gamma z_i, \tag{7}$$

where $d$ is the distance between the capacitor plates and $z_i$ is the perpendicular distance of the ion from the ground plane of the junction capacitor.

From Eqs. (4) and (7) we can now write the energy splitting term due to the i-th ion as:

$$\Delta_i = \frac{4\pi Q_i}{Cd\Phi_0}\langle 0|p_\gamma|1\rangle\langle e_i|z_i|g_i\rangle \ . \tag{8}$$

Using a harmonic approximation for the junction states, where $\hbar\omega$ is the energy level spacing, we find:

$$<0|p_\gamma|1> = -i\frac{\Phi_o}{2\pi}\sqrt{\frac{\hbar C\omega}{2}} \ . \tag{9}$$

For an ion moving in a local two-well potential, the overlap integral term is:

$$<e_i|z_i|g_i> \approx \frac{\delta z_i T_i}{\sqrt{4T_i^2 + U_i^2}} = \frac{\delta z_i T_i}{E_i}, \tag{10}$$

where $\delta z_i$ is the difference in the perpendicular distance between the two wells of the potential, with respect to the ground capacitor place, $T_i$ characterizes the strength of the tunneling between the two wells, and $U_i$ is the difference in the energy of the minima of the two-wells [29]. Substituting Eqs. (9) and (10) into Eq. (8) and setting $E_\gamma = E_i$ at the avoided crossing, we find the energy splitting due to the ion is:

$$\Delta E_i = |\Delta_i| = \frac{\delta z_i}{d}\sqrt{\frac{Q_i^2}{2CE_i}}2T_i \ . \tag{11}$$

If the wells are of the same depth, i.e. $U_i = 0$, then $E_i = \sqrt{4T_i^2 + U_i^2} = 2T_i$, $\langle e_i|z_i|g_i\rangle = \delta z_i/2$ and the energy splitting will be a maximum [30]:

$$\Delta E_i^{max} = |\Delta_i^{max}| = \frac{\delta z_i}{d}\sqrt{\frac{Q_i^2 E_i}{2C}} \ . \tag{12}$$

We can also consider avoided level crossings between higher level states of the system such as $|2g\rangle$, $|1e_i\rangle$, and states such as $|0e_ie_j\rangle$ where two TLSs are excited. Given a model of each TLS, the energy and splitting size of these higher energy states are predictable based on the

lower energy spectrum. We note that states such as $|1e_i\rangle$ need to be considered when analyzing Rabi oscillations in the coupled systems [12, 14] and the $|2g\rangle$ state may be relevant in gate operations [31] and readout [32]. States $|0e_i e_j\rangle$ will occur in the same energy range as $|1e_i\rangle$ and $|2g\rangle$ if $E_\gamma \approx E_i \approx E_j$. For example, the splitting term between $|1e_i\rangle$ and $|2g\rangle$ is:

$$\Delta_i = \frac{4\pi Q_i}{Cd\Phi_0} \langle 1|p_\gamma|2\rangle\langle e_i|z_i|g_i\rangle. \tag{13}$$

Using the harmonic oscillator approximation for the states of the junction, we find $\langle 2|p_\gamma|1\rangle = \sqrt{2}\langle 1|p_\gamma|0\rangle$, and thus the avoided crossing between $|2g\rangle$ and $|1e_i\rangle$ will be 40% larger than the corresponding avoided crossing between $|1g\rangle$ and $|0e_i\rangle$ if all other parameters are identical.

**Splitting due to critical current fluctuators**

For a critical current fluctuator, an interaction arises when the position of an ion affects the critical current of the junction. If we assume that the ion is moving between a left and right well in a symmetric double well potential, then the coupling term between $|1g\rangle$ and $|0e_i\rangle$ can be written as [11, 12]:

$$\Delta_i = 2\langle 0e_i|H_{ci}|1g\rangle = -\frac{\Phi_0 \delta I_0}{2\pi}\langle 0|\cos(\gamma)|1\rangle, \tag{14}$$

where $\delta I_0$ is the difference in the critical current when the particle is in the right and left well of the two-well potential. If we make a harmonic approximation for the junction states, we find the splitting energy can be written:

$$\Delta_i \approx \frac{\Phi_0 \delta I_0}{2\pi} \sqrt{\frac{2e^2}{CE_\gamma} \frac{I}{I_0}}, \tag{15}$$

where $I$ is the current bias applied to the junction. The splitting term for $|2g\rangle$ and $|1e_i\rangle$ is:

$$\Delta_i = 2\langle 1e_i | H_{ci} | 2g \rangle = -\frac{\Phi_0 (\delta I_0)}{2\pi} \langle 1 | \cos(\gamma) | 2 \rangle. \tag{16}$$

If the current at which the two splittings occur are the same, then $<1e_i | H_{ci} | 2g> \approx \sqrt{2} \langle 0e_i | H_{ci} | 1g \rangle$. Since this is the same relationship as for a charge fluctuator a comparison of the size of the $|2g\rangle$ and $|1e_i\rangle$ splitting to the size of the $|1g\rangle$ and $|0e_i\rangle$ will not distinguish the manner in which the fluctuator couples to the junction. However, examination of Eqs. (8) and (14) reveals that the splitting size for a critical current fluctuator depends on the current bias and critical current, while the splitting size for a charge fluctuator does not. Thus in principle, a charge fluctuator can be distinguished from a critical current fluctuator, if splittings caused by the same TLS can be measured at different currents or critical currents.

Finally, we note that in the harmonic oscillator approximation, both the critical current and charge models yield no coupling between $|0e_i\rangle$ states and the $|2g\rangle$ state. However, for the more realistic cubic approximation [31] we find a small coupling in both models:

$$\langle 0e_i | H_{ci} | 2g \rangle = \lambda \langle 0e_i | H_{ci} | 1g \rangle \approx \frac{1}{\sqrt{54 N_s}} \langle 0e_i | H_{ci} | 1g \rangle, \tag{17}$$

where $\lambda$ is the anharmonicity and $N_s$ is the number of levels in the potential well. Under our typical bias conditions, $N_s \sim 5$, $\lambda \sim 0.06$, and thus we expect fluctuators to produce an order of magnitude smaller splitting if they intersect the $n=2$ level of the junction instead of the $n=1$ level. We also note that there is no coupling between the $|1g\rangle$ state and $|0e_i e_j\rangle$ states in our model because we have assumed that the microstates do not interact with each other.

**Experimental arrangements**

Figure 1(a) shows a schematic of a dc SQUID phase qubit. The qubit junction J1 and an inductor $L_1$ are connected across a small inductor $L_2$ and an isolation junction J2 resulting in a dc SQUID. As first pointed out by Martinis *et al.* [32], this configuration provides broadband inductive isolation of the qubit junction J1 from current noise on the bias leads. To isolate the loop from noise in the flux bias, the mutual inductance $M$ between the SQUID and the flux bias coil must be much smaller than the total loop inductance. The current through each junction can be independently controlled by simultaneously ramping the bias current and flux applied to the loop in the proper proportion. Provided $L_1$ is sufficiently large, the qubit junction J1 acts as a single well-isolated Josephson junction phase qubit [27]. The two lowest energy levels in one well of the qubit's washboard potential form the qubit states $n = 0$ and $n = 1$. Higher energy levels will also be present in the well, with an anharmonic level structure. Varying the current through the qubit junction changes the energy difference between the qubit levels. By applying microwaves resonant with the energy level spacing, we can excite transitions and map out the energy levels [34].

Figure 1(b) shows a photograph of device DS3A, which was built on a sapphire substrate and had no $SiO_2$ or other wiring insulating layers deposited. We used photolithography to form a resist bridge and then performed double-angle evaporation of aluminum to create the tunnel junctions [35]; the Al/AlO$_x$/Al qubit tunnel junction and SQUID loop were formed by evaporating from one angle roughly 20 nm of Al in a high vacuum, oxidizing for 10 min in 10 Torr of $O_2$, depositing 30 nm of Al at a second angle, and then lifting off the pattern in photoresist remover [36]. The resulting qubit junction had an area of 16 $\mu m^2$.

During measurements the devices were mounted in a superconducting aluminum shield box that was attached to the mixing chamber of a dilution refrigerator. All leads to the device were heavily filtered and measurements were made at 25 mK. We use a pulsed-readout technique to measure the state of the system. This involves applying a brief current pulse to the qubit junction, via a small capacitor, and checking if the device escapes to a different flux state [1, 37, 38]. The technique exploits the fact that each successive energy level has a tunneling rate that is two to three orders of magnitude faster than the level below it. By carefully choosing the size of the current pulse, one can ensure that the $n=1$ level (and all higher levels) will tunnel with a high probability, while the ground state $n=0$ level will tend not to.

Table 1 summarizes the measured parameters for device DS3A. Testing of the device revealed that we could drive Rabi oscillations in the qubit with Rabi decay times $T'$ of up to 27 ns and relaxation times $T_1$ up to about 28 ns, with both $T'$ and $T_1$ being sensitive to the qubit transition frequency [39]. We also built and tested a second device DS4A which had the same design and similar parameters, except for a larger loop inductance $L_1$.

**Spectroscopic measurements of fluctuators**

Figure 2 shows a gray scale plot of the transition spectrum measured in device DS3A in the frequency range between 10 GHz and 11.2 GHz. For this data, the qubit was biased at approximately 60 different currents, continuous microwaves were applied in 5 MHz or 3 MHz steps and the average population in $n=1$ was measured by repeatedly applying single-amplitude measurement pulses at each bias point. Examination of Fig. 2 reveals that several avoided crossings are clearly visible with sizes ranging up to about 250 MHz. Splittings less than about 10 MHz are probably present, but are not clearly resolved because of the width of the transitions and the finite size frequency steps used in the measurements.

While all of the avoided crossings may not be of identical origin, a comparison with other devices we have tested suggests that they are closely related to the qubit junction. Two other dc SQUID qubits with 16 µm$^2$ area Al/AlO$_x$/Al junctions showed a similar density of prominent splittings. We also found that the frequency at which an avoided crossing occurred was independent of the bias current through the isolation junction.

In contrast, several dc SQUID qubits with 100 µm$^2$ area Nb/AlO$_x$/Nb junctions tested in the same manner, showed what appears to be a large number of splitting over a 1 GHz range, but no splitting greater than 10 MHz [40]. This made it very difficult to do the type of study we describe here, so a detailed comparison is not possible.

To analyze the avoided crossings, at each current bias the probability of escape $p_e$ versus microwave frequency $f$ was $\chi^2$-fit to a sum of Lorentzian peaks, with one Lorentzian for each distinguishable peak. Figure 3 shows four sample fits for different currents through the qubit junction. An offset in $p_e$ of about 4% for frequencies far from a peak is due to population in $|0\rangle$ tunneling during the readout pulse. We note that when the qubit is coupled to a TLS, simply adding two Lorentzians may not be rigorously correct. However, in practice it gives good fits for the spectroscopic width and peak location. Away from any splittings, we found good fits of $p_e$ versus $f$ to a single Lorentzian. The data in Fig. 3 was taken several months after that shown in Fig. 2, yet most of the splittings remained relatively stable.

Figure 4 summarizes the result of the fitting analysis. The circles in Fig. 4 correspond to the location of the resonance peaks and the vertical lines are not error bars but instead represent the FWHM found using the fitting procedure at each current. The circles at 10.714 GHz correspond to "peaks" that were not Lorentzian. The enhancement for these bias currents immediately drops to zero at frequencies above 10.714 GHz, regardless of the bias current or the enhancement at frequencies less than 10.714 GHz. This suggests that microwaves drive was blocked from the qubit in this frequency range. This might be due to a resonance in the sample box, for example, but we did not investigate it further and have ignored these points in further analysis of the spectrum.

Figure 5(a) shows a detailed view of a section of Fig. 4 with labels for several of the states. The solid curves in Figs. 4 and 5 are a fit to the energy levels of the qubit coupled to eight two-level systems. The uncoupled qubit spectrum was found by solving Schrodinger's equation for a single Josephson junction, which depends on only the critical current; here $I_0 = 1.263 \, \mu A$ and capacitance of the junction qubit $C = 0.377 \, pF$. In the fit to the full spectrum, we assumed that the i-th TLS has a first excited state energy $E_i$ and a coupling term $\Delta_i$ that is independent of $I$ and does not directly couple to other microstates. For the fit curves in Figs. 4 and 5 each $E_i$ and $\Delta_i$ and the qubit parameters $I_0$ and $C$ were varied to determine the best fit to the peak locations for the entire spectrum simultaneously. The resulting best-fit values are given in Table 2. Examination of Fig. 4 shows that we obtained excellent agreement between the data and the fit, especially considering the simplicity of the model. While we assume the microstates do not couple directly to each other, the fit is far from trivial because each TLS perturbs the junction and thereby causes an indirect perturbation of other avoided crossings that are nearby in frequency.

**Distribution of the splitting sizes**

For our device parameters, the maximum splitting size in a charge-coupling model with $Q = e$ is approximately 700 MHz at 10.5 GHz (see Eq. 12). A splitting size of this magnitude would be unlikely in our model, since the two wells, which we assume are randomly positioned, would need to be on opposite ends of the junction barrier. The largest splitting we observed was 240 MHz at 10.935 GHz corresponding to a single electron charge hopping a perpendicular distance $z$ of about 1/3 of the dielectric thickness, or about 3 Å for an $AlO_x$ barrier thickness of 1 nm.

Figure 6(a) shows the distribution of the splitting sizes. The points show the cumulative number of splittings with sizes larger than 9 MHz and less than $\Delta_f'$, plotted as a function of $\Delta_f'$. We also measured splittings in device DS4A and found five splittings (16, 17, 27, 94, 115 MHz) over an 800 MHz frequency range; these splitting are included in the data, plotted in Fig 6(a). The curve is a fit to the distribution expected if the splittings are caused by ions that are tunneling between random locations in the tunnel junction [39], with a maximum splitting size $\Delta_{f\,max} = 700\,\text{MHz}$. The observed distribution of splittings is in good qualitative agreement with the model, and the overall number and distribution is comparable to and slightly lower than Martinis *et al.* reported for splittings in their 13 µm² $Al/AlO_x/Al$ phase qubits [2].

**Distribution of TLS spectroscopic coherence times**

From our Lorentzian fits, we also determine the full-width-at-half-maximum (FWHM or $\Delta f$) of the resonant peaks as a function of current (see Fig. 5(b) and 7). In Fig. 7, two different symbols (circles and triangles) are used to clarify distinct branches in the spectrum. The right hand axis in Fig. 7 shows the corresponding spectroscopic coherence time $T_2^*$, where $T_2^* = 1/(\pi \Delta f)$. Figure 5(b) shows an enlarged view of the fit in the same region as Fig. 5(a) to help clarify the relationship between the different branches of the spectrum and the corresponding FWHM. Both figures reveal large changes occur in the FWHM near the avoided crossings.

The solid lines in Fig. 7 and 5(b) are from our model of the coupled system. Imaginary energy terms were added to the qubit energy and eight microstate energies in the Hamiltonian of Eq. (2); then the eigenvalues of the system at each bias point were again determined. The imaginary energy terms were all fit simultaneously to give the best $\chi^2$-fit. We assume that the width of the qubit resonance does not depend on the bias point. The coupling strengths and real part of the energies were completely determined from fitting the peak Lorentzian locations, only the coherence times were varied to fit the widths (see Table 2 for the resulting fit parameters). We generally observe good qualitative agreement, with the $\Delta f$ reflecting its composition from the different quantum systems. As expected, eigenstates that are predominantly an individual TLS have a fixed $\Delta f$ that then changes as the crossing is approached. However, some discrepancies are clearly visible. For example, in Fig. 5(b) at $I \approx 1.122\,\mu A$, the observed crossing in the $\Delta f$ points does not quite match the predicted crossing. Finally, we found that at $I \approx 1.108\,\mu A$, the peaks are very poorly fit by Lorentzians and at $I \approx 1.10\,\mu A$ the data is very scattered, and thus no fits are shown for $I < 1.108\,\mu A$.

Our best fit to the entire spectrum required a qubit $\Delta f = 12\ MHz$ or $T_2^* = 27\ ns$ as shown in Table 2. Experimentally, the best spectroscopic $FWHM$ measured for the qubit at a single bias point was $\Delta f = 15\ MHz$ or $T_2^* = 21\ ns$ with the peak fitting well to a single Lorentzian. At this same bias point Ramsey fringe measurements were observed to decay with a very comparable characteristic time, $T_2^{ramsey} \approx 20-25\ ns$. These values are fairly consistent, but suggest that even when the qubit is operated away from obvious avoided crossings and the spectrum fits well to a single Lorentzian, the TLSs may still be slightly influencing the qubit. We have not been able to definitively explain the coherence time and relaxation rate of the qubit when it was not coupled to any noticeable splittings. The limiting factor could be dielectric loss in the tunnel barrier from splittings too small to observe in the qubit spectrum. We note that as the current through the junction is decreased, $\partial f_{01}/\partial I$ decreases, suggesting a smaller spectroscopic width should be expected at higher frequencies if the device is limited by low-frequency current noise (a similar argument can be made for flux noise or critical current noise) [41]. This effect is not evident in our data.

Examining Table 2, we note that the TLSs show a wide range of line widths and several of the microstates had $\Delta f \leq 6\,MHz$ corresponding to spectroscopic coherence times $T_2^* > 50\,ns$. For example, fine scale measurements with 1 MHz steps in a region where the 4$^{th}$ microstate (at $f_4 = 10.472\,\text{GHz}$) was only very weakly coupled to the qubit showed $\Delta f = 4\,\text{MHz}$ or $T_2^* = 80\,\text{ns}$. Time domain measurements, similar to those performed in Ref. [1], when this microstate was coupled to the qubit also verified it had a longer relaxation time than the qubit. Interestingly, the 3$^{rd}$ and 7$^{th}$ microstates have markedly larger FWHM ($\Delta f = 80\,\text{MHz}$ or $T_2^* = 4\,\text{ns}$) along with markedly larger splitting sizes $\Delta_3 = 114\,\text{MHz}$ and $\Delta_7 = 240\,\text{MHz}$. While these splittings may have a different physical origin or coupling than the others, Fig. 6 suggests they are within the expected distribution of splitting sizes.

Figure 6(b) shows a plot of the cumulative distribution of $T_2^*$ for the microstates in device DS3A. The open circles, which represent $N(\tau)$, are the total number of two level systems with $T_2^*$ greater than $\tau_{\min} = 1\,\text{ns}$ and less than time $\tau$, are plotted as a function of $\tau$. Although there were only eight TLSs measured, the resulting curve is remarkably straight when plotted with a linear y-axis and log scale on the x-axis. In particular, the straight line in this figure shows the expected distribution if the number of microstates with a given $\tau$ scaled inversely with $\tau$. In this case, the cumulative distribution is simply:

$$N(T_2^* < \tau) = \int_{\tau_{min}}^{\tau} \frac{A}{\tau} d\tau = A \ln(\tau) - A \ln(\tau_{min}) \qquad (17)$$

where $A = 1.4$ from our fit. We note that this theoretical distribution is what would result if each $T_2^*$ was limited by the lifetime of the TLS and the two-level systems had uniformly distributed tunnel barriers that set the lifetime of the excited state. If this distribution extended to longer and shorter lifetimes, it would lead to a dielectric loss tangent that was independent of frequency and charge noise with a $1/f$ noise power spectrum [42]. This simple distribution appears to be in reasonable qualitative agreement with our data, which supports the idea that the two-level systems we observe in the spectrum are of the same nature as those causing familiar dielectric loss.

**Spectroscopy of higher levels**

To check the higher-level spectrum, we applied microwaves in the 20.4 GHz -21.25 GHz range to excite $n = 0 \rightarrow n = 2$ qubit transitions. This range was chosen since it should yield $|1e_i\rangle$ to $|2g\rangle$ avoided level crossings from the same TLSs seen in Fig. 4 at lower frequencies as $|0e_i\rangle$ to $|1g\rangle$ avoided level crossings. Figure 8 shows the corresponding gray scale plot of the spectrum. Unlike the $n = 0 \rightarrow n = 1$ spectrum, two "tilted" avoided crossings are visible, while no clear "horizontal" splittings are evident. This range includes frequencies where the second excited level would be if the TLS were actually a simple harmonic oscillator. If the TLS is modeled as a simple harmonic oscillator the coupling would still be negligible for both coupling Hamiltonians described above (see Eqs. 8 and 14). Further measurements did reveal horizontal resonances or peak streaks of unknown origin at 19.93 GHz and 21.42 GHz.

At each bias point, we fit the spectrum to a sum of Lorentzians, with one Lorentzian for each clearly distinguishable peak; often more than two were required. Four sample slices at different fixed current are shown in Fig. 9. While the resolution is worse than for the lower frequency spectrum, due to the larger widths and greater number of peaks involved, the data is fairly well represented by the fits. Figure 10 shows the resulting higher-level transitions extracted from the fits. The circles and "error bars" again show the peak location and FWHM at each bias current. The spectra in Figs. 4 and 10 were taken within a week of each other to enable a direct comparison and to minimize bias calibration drift and possible changes in the splitting sizes and locations.

Examination of Fig. 10 reveals clear slanted avoided level crossings. This qualitative behavior is exactly what one expects for the $|2g\rangle$ state coupling to $|1e_i\rangle$ states. The solid curves show the fit to the slanted avoided crossings using the second excited state of the junction and including coupling between the junction and microstates. With the $|1e_i\rangle$ energy and coupling determined from the lower frequency spectrum, the coupling between $|2g\rangle$ and $|1e_i\rangle$ for each microstate was fixed at $\sqrt{2}\Delta_i$, as described above for harmonic states in the junction. For the fits in Fig. 10, the Hamiltonian must be expanded to include $|1e_i\rangle$ and $|2g\rangle$ states. Unfortunately, a simple single junction model cannot fit both the $n=0 \to n=2$ transition and $n=0 \to n=1$ transition, even away from splittings. This may be because the qubit is not actually a single current bias Josephson junction, which we assumed above, but part of a dc SQUID [43]. To obtain the fit shown, a single junction model with a critical current of $I_0 = 1.266\,\mu A$ and $C = 0.374\,pF$ was used for the $n = 2$ state, slightly different than what we used for the $n = 1$ transition ($I_0 = 1.263\,\mu A$ and $C = 0.377\,pF$). The resulting curves are in good agreement with the data, although some discrepancies are obvious. We note that if we allow the couplings of the three slanted avoided crossings to be free parameters in the $\chi^2$ fits (with the locations still fixed), we find the best coupling fits are 1, 1.5 and 2 times the lower coupling.

**Conclusions**

In conclusion, we measured the multi-level spectrum of eight two-level systems coupled to an Al/AlO$_x$/Al dc SQUID phase qubit with no SiO$_2$ dielectic. The complicated multi-level spectrum was accurately fit by assuming that the two-level systems only coupled to the junction, and not directly to each other. By fitting the lower frequency spectrum, we were able to correctly predict interactions between higher energy states of the system. From the fits we were able to extract the energy levels, coupling strength and spectroscopic coherence times of each TLS. The distribution of splitting sizes was consistent with the qubit coupling to charged ions that were tunneling between random locations in the tunnel junction oxide. The distribution of coherence times was consistent with the tunnel barriers having a uniform distribution of heights, as expected for a material that displays a constant dielectric loss tangent or $1/f$ charge noise.

Finally, we note that previously we made fine spectroscopic measurements on two 100 $\mu m^2$ Nb/AlO$_x$/Nb phase qubits made on a SiO$_2$ substrate [40]. These measurements showed no prominent splittings but numerous possible small splittings, all of which were very difficult to resolve because they were less than 10 MHz in size. These devices showed qubit relaxation times $T_1 \sim 15\,\text{ns}$, likely limited due to two-level systems and dielectric loss in the tunnel barrier or surrounding dielectric. As expected, moving to a 6 times smaller qubit junction area, with no insulating dielectric on a $Al_2O_3$ substrate, reduced the number of splittings while increasing the size of the splittings and this allowed for a more complete analysis of the TLSs. However, the increase in the splitting size from the Nb devices to the Al devices is more than can be accounted for based simply on the expected scaling with the junction capacitance as given by Eq. (12), suggesting that the distribution or nature of the TLSs in the Al and Nb devices are different. This is perhaps surprising because the Nb junctions are actually Nb/Al/AlO$_x$/Nb so that the dielectric is nominally the same. In both cases the AlO$_x$ barrier was thermally grown, but the NB devices used sputtered Al, where as the Al devices were evaporated. This would suggest making measurements on Al devices in which the Al has been sputtered, and in devices with a Nb counter-electrode is important.

We would like to acknowledge informative discussions of decoherence, dielectric loss and two-level systems with K. Osborn, B. Palmer, Z. Kim, R. Simmonds and F. Nori and J. M. Martinis and thank M. Kushner, D. Benson and C. Vlahacos with providing assistance for parts of the experiment. This work was supported by Laboratory for Physical Sciences, the Joint Quantum Institute, and the state of Maryland through the Center for Nanophysics and Advanced Materials.

**Tables**

Table 1: Parameters of dc SQUID phase qubit DS3A. $I_{01}$ is the critical current of the qubit junction, $C_1$ is the capacitance of the qubit junction, and $A$ is the area of the qubit junction. $I_{02}$ is the critical current of the isolation junction, $L_1$ is the inductance of the arm of the SQUID that contains the qubit junction, $L_2$ is the inductance of the arm of the SQUID that contains the isolation junction (measured from SQUID $I-\Phi$ curves), and $M$ is the mutual inductance between the SQUID loop and the flux coil.

| | |
|---|---|
| $I_{01}$ (μA) | 1.26 |
| $C_1$ (pF) | 0.37 |
| $A$ (μm)$^2$ | 16 |
| $I_{02}$ (μA) | 8.5 |
| $L_1$ (nH) | 1.05 |
| $L_2$ (pH) | 20 |
| $M$ (pH) | 1.4 |

Table 2: Best fit microstate parameters obtained by fitting the spectra shown in Figs. 4-7. $f_i$ (GHz) is the transition frequency of the i-th TLS, $\Delta_{fi}$ (MHz) is the size of the splitting at the avoided crossing, $\Delta f$ (MHz) is the full-width-at-half-maximum and $T_2^*$ is the spectroscopic coherence time. The last row gives the best fit values for the FWHM and spectroscopic coherence time for the phase qubit.

| TLS # | $f_i$(GHz) | $\Delta_{fi}$(MHz) splitting | $\Delta f$ (MHz) FWHM | $T_2^*$ (ns) |
|---|---|---|---|---|
| 1 | 10.075 | 14 | 23 | 14 |
| 2 | 10.197 | 16 | 6 | 53 |
| 3 | 10.335 | 114 | 80 | 4 |
| 4 | 10.472 | 36 | 2 | 159 |
| 5 | 10.540 | 24 | 17 | 19 |
| 6 | 10.621 | 24 | 5 | 64 |
| 7 | 10.935 | 240 | ~ 80 | 4 |
| 8 | 11.042 | 60 | < 20 | >15 |
| Qubit | $f_{01}(I_0, C, I)$ | | 12 | 27 |

**Figure Captions**

Fig. 1. (a) Schematic of dc SQUID phase qubit. The qubit junction J1 is isolated from the bias line by a large inductance $L_1$ and a smaller inductance $L_2$ and isolation junction J2. The flux line and current bias line enable independent control of the current through each junction.
(b) Photograph of Al/AlO$_x$/Al dc SQUID phase qubit built on a sapphire substrate.

Fig. 2. Transition frequency spectrum of device DS3A as a function of current $I$ through the qubit junction. The false color map represents probability of escape during a pulse (see inset scale). Several prominent splittings are clearly visible.

Fig. 3. Lorentzian fits to resonant peaks in the spectrum shown in Fig. 2. At each bias current, $p_e$ was $\chi^2$ fit with an appropriate number of Lorentzians. From these fits the location and full-width-half-max $\Delta f$ for each peak was determined.

Fig. 4. Transition spectrum showing the peak location (circles) and FWHM (vertical lines) from fitting each bias point. The solid line is a fit to the data using a single junction model, and 8 two-level systems. The frequency and coupling strength for each TLS was varied to give the best $\chi^2$ fit. Each TLS is assumed to only couple to the junction and not directly to other TLSs. Table 2 shows the best-fit values used for this fit.

Fig. 5. Enlarged sections of Figs. 4 and 7. (a) An enlarged section of the transition spectrum in Fig. 4 with several states labeled. Circles and triangles are used to clarify the different branches. (b) Corresponding enlarged section of the FWHM versus current from Fig. 7 with the same

states. The dashed curve is the fit to the triangle points and the solid curves fits the circles. The $\Delta f$ of the junction as determined from the entire fit is plotted as a horizontal line at $\Delta f = 12\,MHz$.

Fig. 6 (a). Number of splittings per GHz with splitting size smaller than $\Delta f'$ and at least 10 MHz versus splitting size $\Delta f'$. The curve is a fit to theory with $\Delta f_{max} = 700\,MHz$ and the total number of splittings as a free parameter. (b) Points show measured cumulative distribution $N(\tau > T_2^*)$, the number of microstates with spectroscopic coherence times $T_2^*$ less than time $\tau$, plotted as a function of $\tau$. Line shows a fit to expected distribution corresponding to a frequency-indepedent loss tangent in the dielectric.

Fig. 7. Full-width-at-half-max and spectroscopic coherence times $T_2^*$ of the microstates and qubit as a function of current through the qubit. The circles and triangles represent the measured $\Delta f$ found from Lorentzian fits. Different shapes are used to help clarify the branches. With the microstate energies and couplings fixed from the fit in Fig. 4, the microstate and junction spectroscopic coherence times were then $\chi^2$ fit (solid curves for the circles and dashed curves for the triangles). When the microstates frequency is significantly far away from the qubit's the fit curves are removed in the plot for clarity.

Fig. 8. Transition spectrum of the qubit and microstates in a higher frequency range, from 20.4 to 21.3 GHz. The false color map represents the fraction that escaped during a measurement pulse. Two "tilted" splittings are visible, one near 20.9 GHz and one near 21.1 GHz. No horizontal

splittings are visible; the light horizontal band near 20.78 GHz is likely due to an anti-resonance in the microwave line, since no bending of the spectrum occurs as the frequency is approached.

Fig. 9. Sample plots of escape events $p_e$ vs. frequency for the higher frequency spectrum. (a-d) Sample slices at fixed current show how several Lorentzians (dashed curves) were combined (solid curve) to fit the measured data points.

Fig. 10. Fit to the higher frequency spectrum shown in Fig. 8. The circles correspond to the fit peaks and the "error bars" are the full-width-at-half-max at each bias point. The solid curves show fits to the second excited state energy of the junction and eight TLSs. The microstate frequencies and coupling strengths are fixed from the lower spectrum with $\langle 2g | H_c | 1e_i \rangle = \sqrt{2} \langle 1g | H_c | 0e_i \rangle$.

(a)

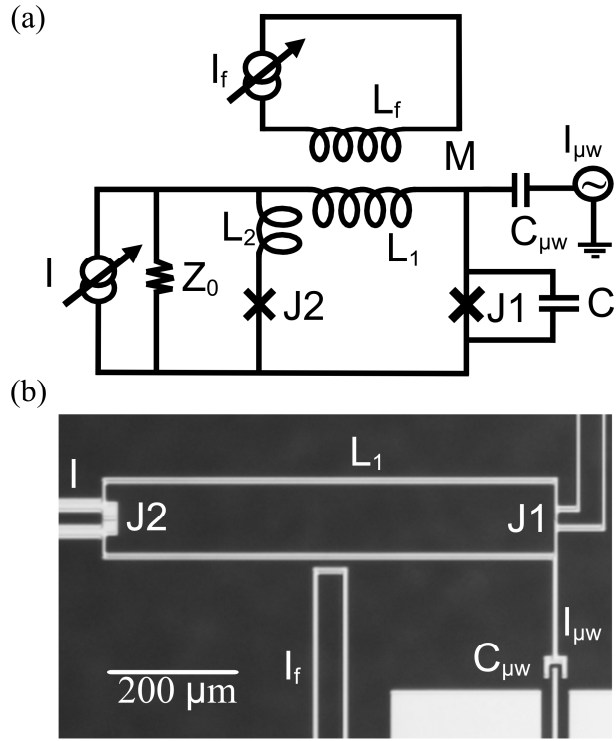

(b)

Fig. 1, Palomaki *et al.*

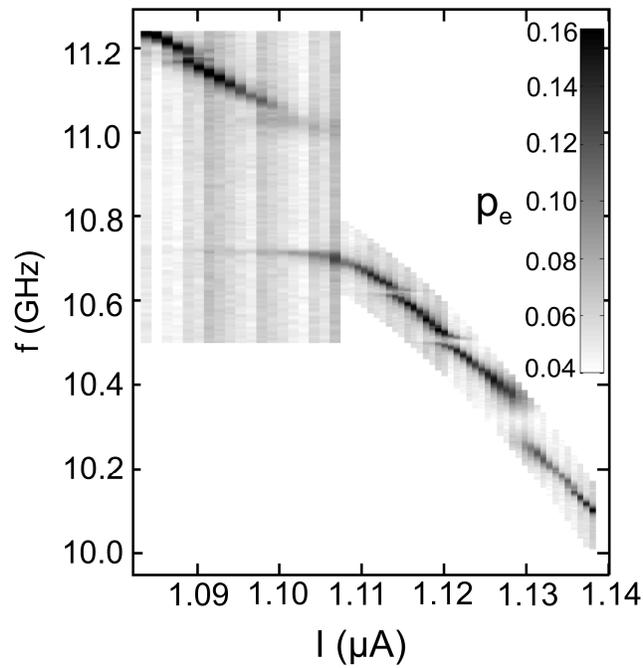

Fig. 2. Palomaki *et al.*

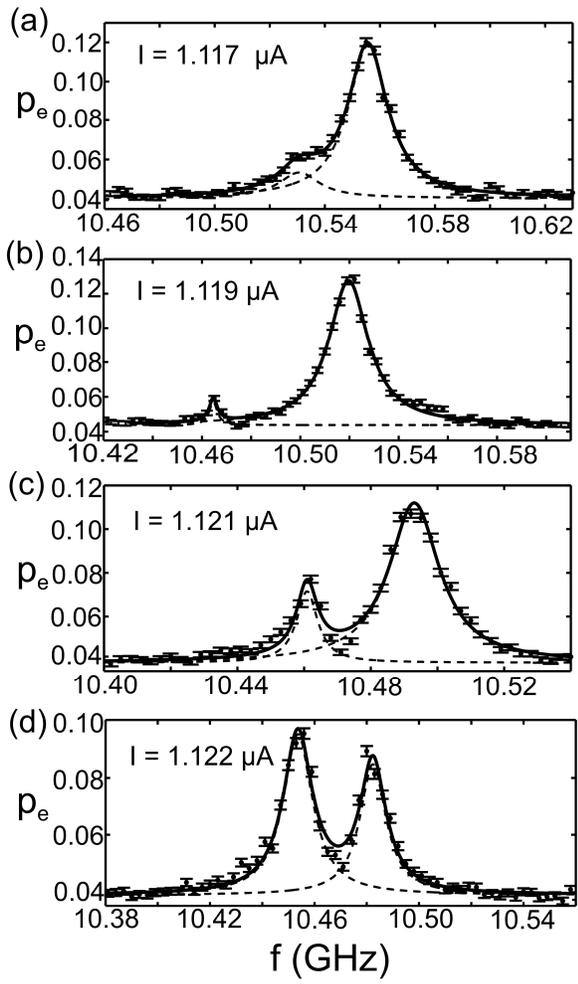

Fig. 3, Palomaki *et al*.

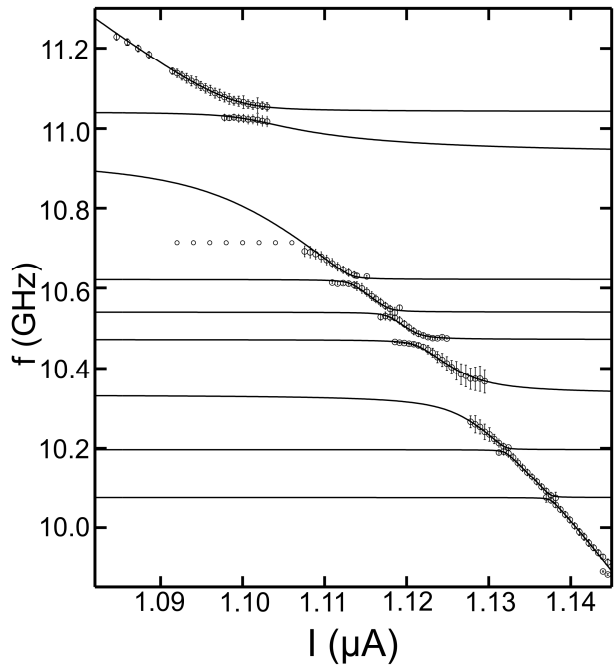

Fig. 4, Palomaki *et al*.

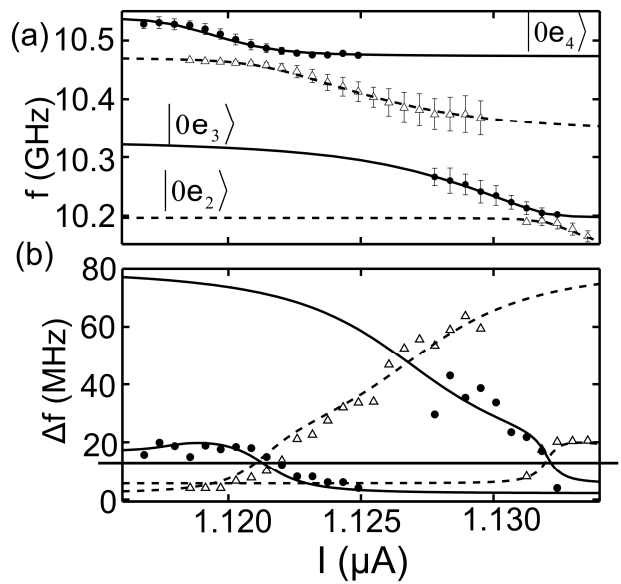

Fig. 5, Palomaki *et al*.

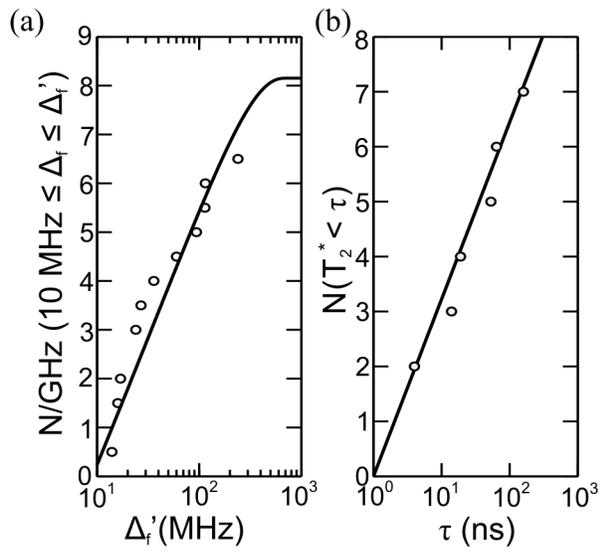

Fig. 6, Palomaki *et al*.

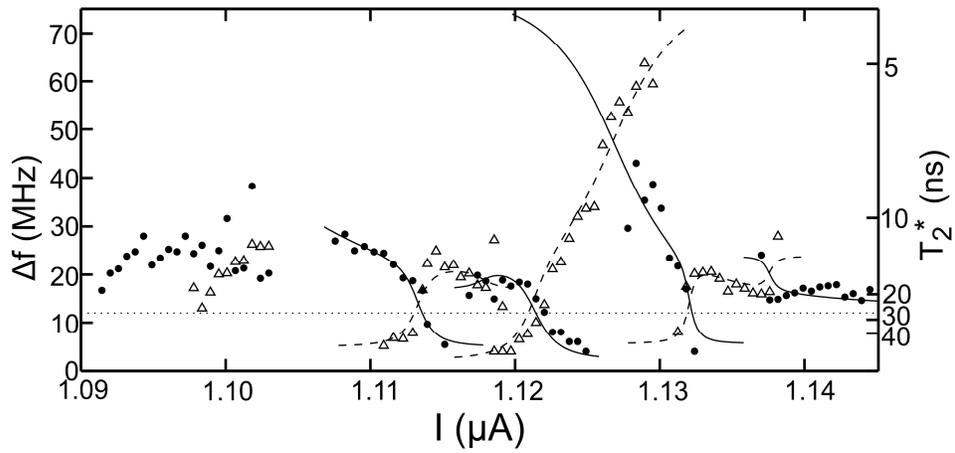

Fig. 7, Palomaki *et al*.

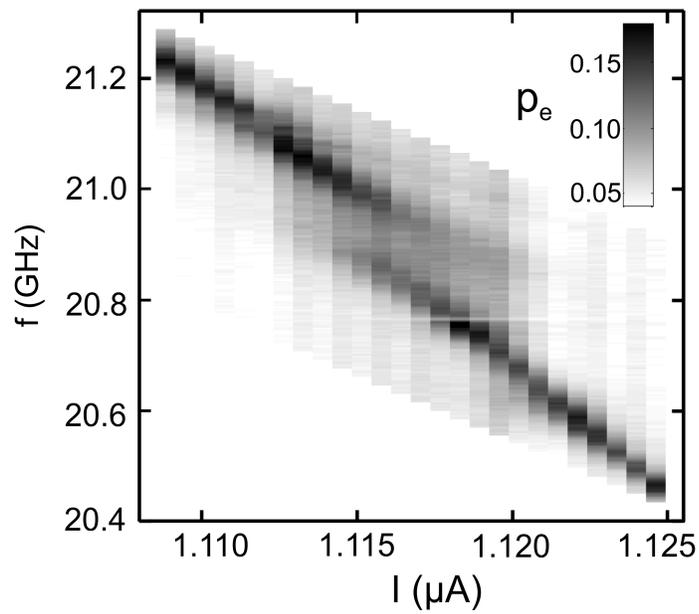

Fig. 8, Palomaki *et al.*

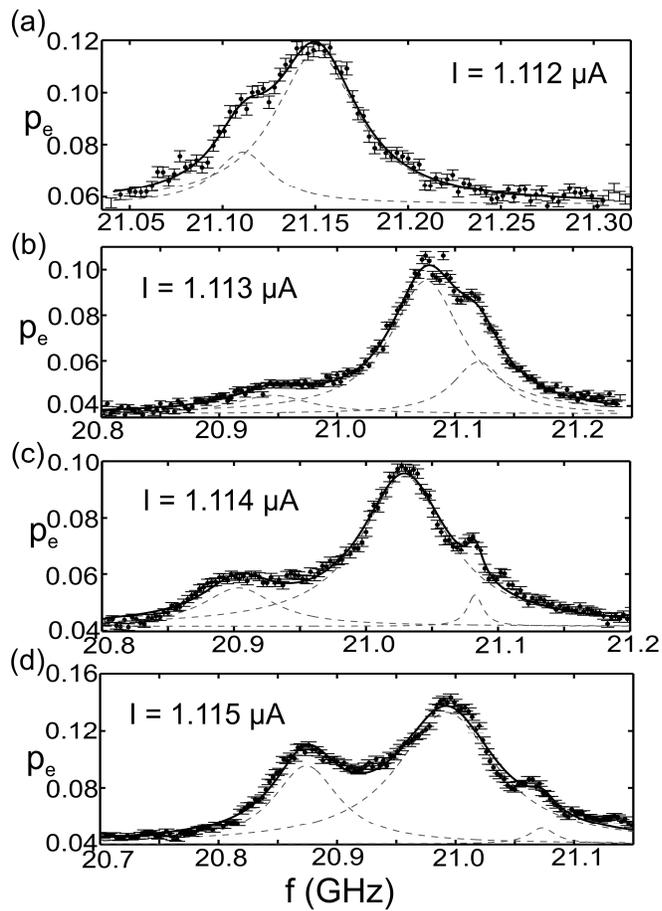

Fig. 9, Palomaki *et al.*

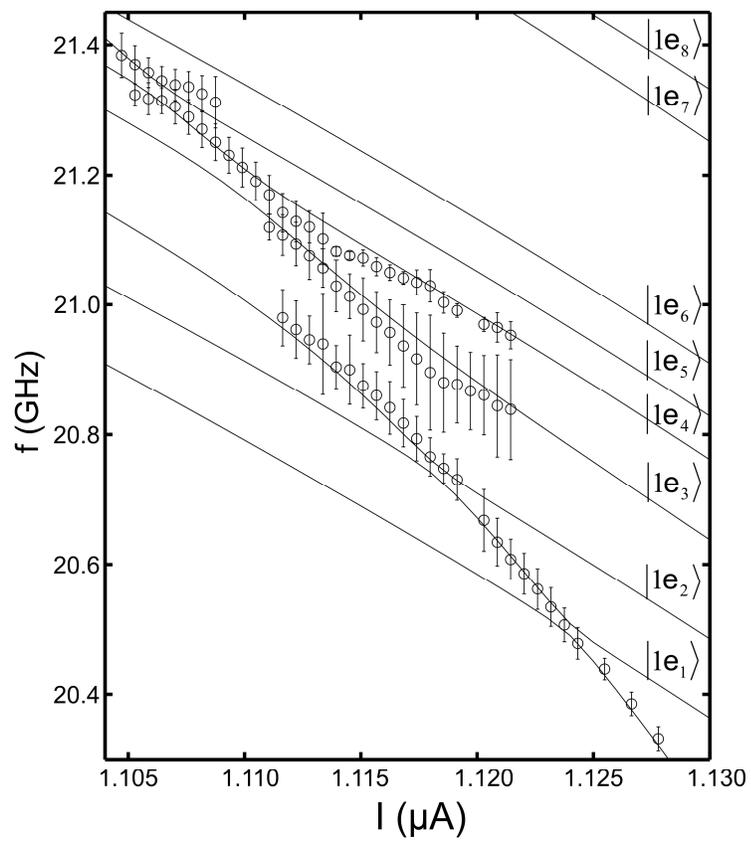

Fig. 10, Palomaki *et al.*